\documentstyle[12pt,amssymbols]{article}
\input math_macros.tex

\begin{document}
\begin{titlepage}
\begin{center}
\today     \hfill    LBL-37315 \\

\vskip .5in

{\large \bf Values and the Quantum Conception of Man}
\footnote{This work was supported by the Director, Office of Energy
Research, Office of High Energy and Nuclear Physics, Division of High
Energy Physics of the U.S. Department of Energy under Contract
DE-AC03-76SF00098.}

\vskip .5in
Henry P. Stapp\\
{\em Lawrence Berkeley Laboratory\\
      University of California\\
    Berkeley, California 94720}
\end{center}

\vskip .5in

\begin{abstract}

Classical   mechanics is  based upon a   mechanical  picture of  nature that is
fundamentally incorrect. It has been replaced at the basic level by a radically
different theory: quantum  mechanics. This change  entails an enormous shift in
our basic  conception of nature, one  that can profoundly  alter the scientific
image  of man   himself.   Self-image is  the   foundation of  values,  and the
replacement of the  mechanistic self-image derived  from classical mechanics by
one concordant  with quantum  mechanics may  provide the  foundation of a moral
order better suited to  our times, a self-image  that endows human life with
meaning, responsibility, and a deeper linkage to nature as a whole.

\end{abstract}
\medskip
\begin{center}
Invited contribution to the UNESCO sponsored Symposium:\\
Science and Culture: A Common Path for the Future\\
Tokyo, September 10-15, 1995
\end{center}
\end{titlepage}

%THIS PAGE (PAGE ii) CONTAINS THE LBL DISCLAIMER
%TEXT SHOULD BEGIN ON NEXT PAGE (PAGE 1)
\renewcommand{\thepage}{\roman{page}}
\setcounter{page}{2}
\mbox{ }

\vskip 1in

\begin{center}
{\bf Disclaimer}
\end{center}

\vskip .2in

\begin{scriptsize}
\begin{quotation}
This document was prepared as an account of work sponsored by the United
States Government. While this document is believed to contain correct
 information, neither the United States Government nor any agency
thereof, nor The Regents of the University of California, nor any of their
employees, makes any warranty, express or implied, or assumes any legal
liability or responsibility for the accuracy, completeness, or usefulness
of any information, apparatus, product, or process disclosed, or represents
that its use would not infringe privately owned rights.  Reference herein
to any specific commercial products process, or service by its trade name,
trademark, manufacturer, or otherwise, does not necessarily constitute or
imply its endorsement, recommendation, or favoring by the United States
Government or any agency thereof, or The Regents of the University of
California.  The views and opinions of authors expressed herein do not
necessarily state or reflect those of the United States Government or any
agency thereof or The Regents of the University of California and shall
not be used for advertising or product endorsement purposes.
\end{quotation}
\end{scriptsize}

\vskip 2in

\begin{center}
\begin{small}
{\it Lawrence Berkeley Laboratory is an equal opportunity employer.}
\end{small}
\end{center}

\newpage
\renewcommand{\thepage}{\arabic{page}}
\setcounter{page}{1}
%THIS IS PAGE 1 (INSERT TEXT OF REPORT HERE)

\noindent{\bf 1. Introduction}

Science has  enriched our  lives in many ways.  It has  lightened the burden of
dreary tasks and  enhanced our  creative capacities. It  has conquered diseases
and extended our  productive years. It  has broadened our  understanding of the
universe   about us  and our  place  within  it. Yet,  while   conferring these
benefits, it has created  the problems of crowding,  pollution, alienation, and
even the threat of  self-extinction. To resolve  these problems a moral base is
needed.   However,  science has  also  largely  destroyed,  at least  among the
educated, the traditional foundation  of morality, namely ancient beliefs about
our link to the  power that created  both ourselves and  the world about us. In
particular,  classical  mechanics, which for  centuries was  our basic science,
transformed  the impulse that  forms and  sustains the world  into a primordial
burst of  energy that set the  universe in  motion, but then  lapsed into total
passivity.  Each man  became, in this  classical  conception, a  mechanical and
microscopically  controlled automata whose every  action was preordained before
he was born. Gone, or  diminished, is the idea that  we bear responsibility for
our actions, for we were taught by  science to see ourselves not as agents of a
creative power, free to choose from among options, but rather as mechanical
devices running on automatic, ruled by forces beyond our control. Science,
having thus undermined the traditional  foundation of morality, seemed to offer
no adequate replacement.

In its  original  seventeenth-century form  classical  mechanics did not wholly
eliminate  the capacity  of spirit and  mind to  influence the  course of human
actions. Thoughts  were allowed to  interact with brains  and, through them, to
affect the motions of our  bodies. But by the  beginning of the present century
both thoughts and gods alike had, according to science, been rendered impotent:
they could do no more than passively  observe the mechanically generated course
of  physical events.  The  clarity and  consistency of  this  conception of the
universe seemed so perfect, and the power of the idea to produce both beguiling
new products  and stable  nations  seemed so  strong, that its  survival seemed
assured. Yet  these concepts  are  fundamentally incorrect.  They are unable to
account for the detailed behavior of  various materials, and by the 1930's this
mechanical conception of  nature had been replaced  at the fundamental level by
something profoundly different: quantum mechanics.

The enormous conceptual gulf between  quantum mechanics and classical mechanics
has  blocked the  dissemination  of this  radically new   conception of man and
nature into the  intellectual  community at large. Hence  its impact upon moral
philosophy has been virtually nil. Yet  one can scarcely imagine that the world
view that  had served  as the  ideological  basis of  the  industrial and early
scientific age can become so thoroughly repudiated without its explosive impact
on our   conception of  ourselves   eventually  asserting  itself.  Indeed, the
greatest remaining gift of science to man may be not a still greater mastery of
our physical  environment, but rather  an unraveling of  the mystery of our own
beingness, and the  consequent rise of  a rational system  of values based on a
more valid self image.

In this contribution to the symposium I shall describe what appears to me to be
the impact  upon moral  issues of the  quantum  revolution in  science. Because
these  questions   appeared to  have no  immediate   professional  relevance to
scientists, the issues have not yet  been widely discussed by those best equipt
to understand them. I shall therefore  endeavour to describe the situation in a
way that  will be  clear to   nonscientists, who  will  need to see  beyond the
technicalities, and also to physicist,  who will want to see, in some form, the
technical basis.

\noindent {\bf 2. From Atom to Man}

Quantum mechanics was  originally a theory about  atoms and their constituents:
it was about our  observations on  systems composed of  electrons, photons, and
atomic nucleii. However, these are the  same elements from which most materials
are made, including the tissues and  other components of our brains and bodies.
Consequently, quantum  mechanics is not merely a  theory about atoms: it is our
fundamental physical theory about the detailed behavior of all material things,
including our own bodies and brains. Yet the  relationship of quantum mechanics
to man goes  far beyond  the fact that  our bodies  and brains  are composed of
atoms. In order to  construct a rationally coherent  theory of atomic phenomena
Niels  Bohr  found it  necessary  to bring  human  observers  into  the theory:
classically   describable   perceptions of  human  observers  became  the basic
realities of the  theory, and the  mathematical formalism  was construed not as
the  description of  the actual  form or  structure of  an  externally existing
reality, but rather as a scheme that  scientist and engineers could use to make
predictions about the structure of their experiences pertaining to a world that
was given no definite  actual form independently of  our experience of it. This
radical  move  was   fiercely  opposed by   Einstein,  and many  other  eminent
physicists   of  that  time.  But they  could  come  up  with no   satisfactory
alternative.

The issue was subsequently re-opened,  and logically acceptable alternatives to
the Bohr interpretation are now available. But the fact remains that any theory
that fits  the empirical  facts must  accept as  elements either
perceptions of human obervers, or  other elements that, like human perceptions,
link together sequences of classically  describable states as {\it alternative}
possibilities,  even  though the basic  quantum  mechanical law  of motion, the
Schroedinger equation, generates no such {\it either-or} decomposition.

There is no  empirical  evidence supporting  the notion that  there is anything
other  than   consciousness, or  mind,  that makes  this  separation  into {\it
alternative}  possibilities, and  chooses between them.  Moreover, if something
else is brought in to do the job, then it is a `stand in' for consciousness, in
the sense that consciousness is all that is needed; and if something else plays
this role, then a  mystery is  generated: Why does  consciousness exist at all?
For if mind does not effect the choices that are needed to complete the quantum
theoretical conception of  nature, then thoughts  appear to have no function at
all in nature: they become superfluous.

Bohr  adopted a very   parsimonious  position: he  brought in only  the minimum
structure needed  to fit the empirical  facts. He  introduced no extra physical
paraphernalia to  define the  alternatives and choose  between them. He let our
perceptions   themselves specify  what has  happened.  The  introduction of our
perceptions  of  the  physical world  into the  basic  physical  theory, though
considered unorthodox during the twenties, can hardly be deemed irrational. For
scientists  rarely deny  the existence  of our  perceptions of  the world. Bohr
merely introduced into  our basic scientific theory  something already known to
exist, and, in fact, the very thing  whose existence is most certain to us, and
whose structure is precisely thing that our science needs in the end to
explain.

Yet Bohr's move  seemed retrograde at  the time. For the  tremendous success of
science was  widely perceived  to be a  vindication of the  wisdom of excluding
spirit and  mind from our  scientific  conception of the  physical world, along
with religious dogmas and myths.

Bohr proceeded very  cautiously with the  re-introduction of mind into science.
Keeping the  connection to the actual  practices of  physicists in the fore, he
and his colleagues, principally Heisenberg, Pauli, and Born, formulated quantum
theory as a set of rules that allowed scientists to calculate the probabilities
that  perceptions conforming  to classically  describable  specifications would
occur under  conditions of this same kind.

Complications pertaining to the living  tissues in the bodies and brains of the
human  observers were kept out  of the theory  by focussing on  the classically
describable   specifications  themselves,  without  worrying about  how we know
whether or not  these conditions are  actually met in real  cases. However, the
pragmatic approach rests  squarely upon our being  able to decide, in practice,
whether such specifications are met or not.

Bohr  could not  evade this  reference to  our  perceptions by  postulating the
existence  of  some  other  classical  level  of  beingness.  For to  admit the
existence of  some other  level of  reality would  contradict  his basic claim,
which was that quantum theory, in the form he proposed, was complete. Admitting
the existence of a  classical level of  physical reality  would require a whole
new  level  of   theoretical   machinery.  This  he   avoided by   allowing our
perceptions, already known to exist,  to be the things that were the subject of
his classically describable specifications.

Although this pragmatic Copenhagen  approach was efficient and practical in the
domain of atomic physics, it provided no detailed idea of how nature managed to
make  the   quantum  rules  work.  This  lacuna  was  of no   great  concern to
practical-minded atomic scientists, but it hindered efforts to extend the scope
of the theory to other domains, such  as cosmology and biology. Heisenberg, von
Neumann, and others  improved the theory in this  respect by providing a theory
for how nature  could work in a way  that would make the  empirically validated
rules come out true.

The key element of  this ontology was  the concept of  `events'. Although there
were  differences among  various  authors regarding  fine  points, the simplest
formulation of the idea  is that the probability  wave of the earlier pragmatic
interpretation, which evolves in accordance with a fixed deterministic equation
of motion, the  Schroedinger equation, is elevated  in status from a subjective
entity  that  scientists  use to  compute   probabilities  pertaining  to their
classically describable  perceptions of the world,  to an objective property of
nature herself.  This objective  property is tied to the  idea of `events': the
probability wave is considered to  define an {\it objective tendency}  for an
{\it  actual event}  to occur.  The  occurrence of any  such  actual event will
reduce  some of  the  uncertainties  that had  existed in  nature  prior to the
occurrence of  this event, and  this reduction  in these  uncertainties will be
reflected in {\it  a new set} of  objective tendencies for  the next event, and
hence a sudden  change in the  probability wave. The fact  that the probability
wave specifies  only  `objective tendencies'  for the next  event, not definite
certainties,   means that  the  particular event  that  will occur  next is not
uniquely determined beforehand: the choice from among the allowed possibilities
is a random event,  with the  statistical weights of the  various possibilities
being specified by the probability wave.

This model  of nature  can be set up  so as to  retreat again  from the idea of
bringing mind into physical theory. That was Heisenberg's tack. But this brings
up the same  problem as  before: it  leaves mind  with nothing  to do. However,
there is no rational  reason to exclude from  physical theory something that we
know exists,  and that seems  to do something,  and then to  bring in, instead,
something else, unknown to us, to do  exactly what the known thing seems to do,
merely because in an earlier {\it and now deposed} theory the known thing could
not do  what  it  seemed to  do,  namely make  real  choices  between  open and
available possibilities.

Von Neumann brought the brains of the observers explicitly into the description
of nature, and stressed  the possibility of  identifying the `choosing events',
needed by quantum theory, with those  brain events that can be considered to be
representations,   within  quantum  mechanically   described brains,  of mental
events. This approach  constitutes, essentially, an  ontological version of the
Bohr  approach,  in that  the  mental  events,  which are  what  specifies what
actually  happens,  are tied  directly  to the  quantum  formalism  without the
explicit introduction of any intermediate classical level of reality.

This von Neumann approach is not the  only ontological possibility. But it can,
I  believe,  be  rightfully   regarded as  the  most  orthodox  of the  quantum
ontologies, for  two reasons. The  first is that it is the  ontology closest in
spirit to  Bohr's  approach: no extra  classical  level  intervenes between the
quantum level of  description and the classically  describable perceptions, and
no profusion of extra unobserved worlds is brought in. The idea that one should
introduce into  physics unverifiable  classical levels of  physical reality is
exactly the  idea that Bohr  fought so  strongly against. The  second reason is
that when the  other quantum  ontologies are considered,  their predictions are
considered  unorthodox to the  extent that the  extra  structure they introduce
produces a  deviation  from the  predictions  obtained without  introducing the
extra structure. This von  Neumann ontology is the  one that leaves out all the
excess structure.

I attribute this ontolgy to von Neumann because his close friend and colleague,
Eugene  Wigner  did so  in a  later  work, in  which  he  extolls and   further
describes  it. Von  Neumann  (1932)  describes this  ontology  briefly, but his
definite preference for it is not  clearly spelled out in his own work. Perhaps
this  approach would  be better  called the  von  Neumann-Wigner  ontology, but
Wigner later rejected it, for reasons that I deem insufficient.

Yet what has all  this discussion  about man and nature to  do with values? The
answer lies in the central importance  to moral philosophy of our beliefs about
such things.

{\bf 3. The Importance of Beliefs}

If a person  truly believes  that doing some  act will cause  him to suffer the
flames of eternal damnation, then he  will probably be disinclined to do it. If
he has no such belief, but believes  himself to be a rotten worthless being who
acts only to benefit himself, regardless of the consequences to others, then he
will probably act in this  way  and thereby become  what he believes himself to
be. If, on the other  hand, he believes himself to  be made of finer stuff, and
the product of a worthy  lineage of high-minded  souls, then he may be inclined
to measure  up to lofty ideals, and thereby to extend the  lineage. What one
believes about himself, and his  connection to the rest of the universe, exerts
a powerful influence on one's behaviour, and it is the whole basis for rational
action.

Science  is a  principal source  of  rationally held  beliefs. If  one believes
himself to be  a mechanically  generated  product of his  genetic make-up and a
mechanically  pre-determined physical environment  then he probably will be far
less able to release his full creative  energy than if he believe himself to be
a facet of a universal impulse in nature that exploits the indeterminateness of
the physical world to actualize intentions and generate meaning. Moreover, from
a rationally based  perception of a  deep-seated wholeness  of nature there can
flow both more compassion and less alienation.

{\bf 4. The Nature of Man.}

What is the quantum mechanical conception of the nature of man?

By the quantum mechanical conception I shall mean, for the reasons given above,
the von  Neumann  conception. I  have in my  book and  elsewhere  (Stapp, 1993,
1995a-c) filled in some  of the details of this  conception in a way that seems
both natural and  compatible with the empirical  evidence from neuroscience and
psychology. The key point is that each  human conscious event is represented in
this conception of nature  by a quantum event that  actualizes {\it an extended
structure} in  the brain of  some human being.  This event  selects, and brings
into being,  one template  for action  from among  many that,  according to the
quantum mechanical laws, were all physically possible just prior to that event.
Each such template is a coordinated  plan of action for this brain and the body
it controls.

In any physical theory of man a primary job of man's brain must be to form such
templates  for  action. The  essential   difference  between the  classical and
quantum  conceptions is that  in the classical  conception the  brain must come
up---quickly in an  emergency  situation---with exactly  {\it one} template for
action, which will direct the unfolding of some coherent action, whereas in the
quantum case, because of Heisenberg's indeterminacy principle, the evolution in
accordance with the  Schroedinger equation will  generate a host of alternative
possible templates for action. Thus if  a situation calling for action presents
itself to  an alert  person, his  brain will  generate  {\it one}  template for
action, according to the  classical conception of  nature, but many alternative
possible templates for  action according to the  quantum conception. It is this
profusion of  possible  templates for action,  and consequent  actions, that is
resolved in the von  Neumann ontology by the  occurrence of an ``event'', which
selects one of the possibilities and eliminates all the others. This event is a
mental event that  is represented in  the quantum  mechanical conception of the
physical world by a  sudden change in the form of  the probability wave, namely
by a jump to a form that has all of  the probability concentrated on the branch
of the  probability wave  that  represents this  chosen course  of action, and,
correspondingly,   a  null   probability  assigned to  all of  the  alternative
possible branches. The actualized template for action is an extended physical
structure in  the brain,  and it is  supposed to  embody all of  the structural
information  that is  contained in the mental event.  Thus the  mental and
physical
events  can be   considered to be  two  aspects of  the same  thing. Each event
represents from the  physical perspective provided  by quantum mechanics a bona
fide free choice from  among open and  available options.

{\bf 5. Chance, Choice, and Meaning}

This quantum conception of man breaks  the bondage of an iron-handed mechanical
determinism.  Man becomes  an aspect  of the  process by which  nature uses the
latitude, or freedom,  expressed by the Heisenberg  indeterminancy principle to
inject form  and structure  into the universe.  In the  classical conception of
nature all freedom to choose was  concentrated at the moment of the creation of
the universe, and hence none was reserved for later use. But quantum theory
transferes this  freedom to later  times, and von  Neumann's conception  shifts
some of it to our thoughts: our minds  become endowed with some of the power to
act freely that in classical mechanics was the prerogative of God alone.

Our  choices  are  not  reclused  from  meaning.   Each   choice is the
expression of an  intentionality. It arises within  a context, and it initiates
an action designed to promote certain  values. The intention of the action and
values it serves are integral parts of the felt act  of choosing.

These qualities of the quantum event can be contrasted with the meaninglessness
of random events that might be imagined to occur at some  microscopic level.
There it is  impossible to embody in  the physical  structure actualized by the
event  any   representation of   intentionality or  value  that  transcends the
momentary   situation. But  the events  of the  von  Neumann  conception, which
actualize extended  physical  structures that are imbedded  in the interpretive
mechanism provided by the brain and body, do embody intentionality, values, and
meaning, all of which are felt at the mental pole.

A healthy brain is designed and  conditioned to produce the actions most likely
to serve the needs and values of the  person, as judged from the perspective of
that person. Of  course, there are  always uncertainties  in our assessment the
physical  situation, and  fluctuations in the  biological  computing machinery.
Hence different  parallel brain  calculations of the best  course of action can
come up with  different  conclusions.  In the  quantum ontology  these parallel
computations are all performed  simultaneously, and the various options are all
presented. The  statistical weight  assigned to each  option is essentially the
number  of  parallel  classical   computations that  lead to  that  option. The
simultaneous availability  of all the options can  be regarded as an expression
of the  freedom that is  represented  by the  quantum  indeterminateness of the
physical  situation. This  indeterminacy makes  the quantum  choice a bona fide
free choice, yet a choice that has  only the latitude allowed by the underlying
physical indeterminacy. The choice is  thus at the same time both a free choice
and yet,  statistically speaking, in  terms of the entire  ensemble of weighted
possible  choices, also the  unique best  choice: this  ensemble is roughly the
statistical ensemble of  computed best actions,  given the indeterminateness of
both the external situation and the internal computational machinery.

These choices are not blind choices,  as they would be if they occurred at the
microscopic level. For  they are choices between  options that project into the
future   actions that  embody   intentions  based on  our  values. The  choices
constitute value-laden intentions, and are thus endowed with meaning.

This image of man is incomparably more inspiring and liberating than the dreary
picture painted by classical  mechanics. Man regains, within limits, control of
his destiny. He becomes  an integral part of  nature's process of infusing
structure and meaning into the universe.  He is granted a portion of the power
that classical mechanics reserved for God alone.

Beyond  its    re-instatement of   personal  freedom and   meaning the  quantum
conception unveils a still deeper  truth. This arises from an aspect of quantum
mechanics not yet touched upon here,  namely the deep-level of connectedness of
spatially separated  physical entities. Once two  entities have interacted they
become intrinsically intertwined in a  way that is not physically apparent, and
that moreover  defies comprehension  within the way of  thinking that underlies
classical  mechanics and our  common-sense  understanding of  nature. Yet it is
entailed by quantum  mechanics, and has been  confirmed by delicate experiments
in simple cases where  sufficient control over the  experimental conditions can
be maintained. This deep-level connectedness entails that our choices, although
highly  personal in  terms of  their meaning  to us,  have another  aspect that
transcends  the   individual. A  choice  made by  one person   generally has an
`instantaneous  effect' on the  objective  tendencies  associated with far-away
entities with whom he has interacted  at some time in the past. It is as if the
entire  universe  is, in  some  sense, a  single  organism  whose  parts are in
instantaneous  communication. This means that  although each of us participates
in an  individually meaningful  way in the  process that  infuses form into the
universe,  and can  shape  this  process in  accordance  with his  own personal
values, nevertheless the  process is basically one  universal activity of which
each of us is a  highly integrated  part. Quantum theory  indicates that we are
all, far more  intricately than  appearances indicate,  facets of one universal
process. Thus, according  to the quantum conception  of nature, the notion that
any one of us is separate and distinct from the rest of us is an illusion based
on  misleading  appearances.  Recognition of  this deep  unity of  nature makes
rational the belief that to act against another is to act against oneself.

{\bf References}

Stapp, H.P. (1993) Mind, Matter, and Quantum Mechanics,\\
    Springer-Verlag, Heidelberg Berlin New York London Paris Tokyo\\ Hong Kong
    Barcelona Budapest

Stapp, H.P. (1995a) {\it Why Classical Mechanics Cannot Naturally\\
    Accomodate Consciousness But Quantum Mechanics Can,} \ PSYCHE 2(6)\\
    ftp://psyche.cs.monash.edu.au/psyche/ psyche-95-2-1-QM\_stapp-1-stapp.txt\\
    http://psyche@cs.monash.edu.au/psyche/public/volume2-1/\\
    psyche-95-2-1-QM\_stapp-1-stapp.html

Stapp, H.P. (1995b) {\it Quantum Mechanical Coherence, Resonance, and Mind,}
    in Norbert Wiener Centenary Congress, V. Mandrakar and P.R.Masini, eds.
    Amer. Math. Soc. Ser. Proc. Sympos. Appl. Math. (PSAPM);\\
    ftp://Phil-Preprints.L.Chiba-U.ac.jp/pub/preprints/Phil\_of\_Mind/\\
    Stapp.Quantum\_Mechanical\_Coherence\_Resonance\_and\_Mind/
    36915.tex

Stapp, H.P. (1995c) {\it The Hard Problem: A Quantum Approach,}
    \ J. Consc. Stud.\\
    ftp://Phil-Preprints.L.Chiba-U.ac.jp/pub/preprints/Phil\_of\_Mind/\\
    Stapp.The\_Hard\_Problem-A\_Quantum\_Approach/37163.txt

Von Neumann, J. (1932)  Mathematical Foundations of Quantum Mechanics,
    Princeton University Press, Princeton N.J. 1955 (English edition)

Wigner, E. (1961) {\it Remarks on the Mind-Body Problem,} in  The Scientist
    Speculates, I.J. Good, ed. Heineman, London

\end{document}